\documentclass[10pt,conference]{IEEEtran}

\IEEEoverridecommandlockouts

\usepackage{cite}
\usepackage{amsmath,amssymb,amsfonts}
\usepackage{algorithmic}
\usepackage{graphicx}
\usepackage{textcomp}

\usepackage{algorithm}

\usepackage{xcolor}
\usepackage[left=0.68in, right=0.673in, top=0.78in, bottom=1.1in]{geometry}

\begin{document}
\title{Test Code Generation for Telecom Software Systems using Two-Stage Generative Model}

\author{\IEEEauthorblockN{Mohamad Nabeel\,, 
Doumitrou Daniil Nimara\,, 
Tahar Zanouda }\
\IEEEauthorblockA{Global AI Accelerator, Ericsson, Sweden}\
\texttt{\{mohamad.nabeel, doumitrou.nimara, tahar.zanouda\}@ericsson.com}}


\maketitle


\begin{abstract}
In recent years, the evolution of Telecom towards achieving intelligent, autonomous, and open networks has led to an increasingly complex Telecom Software system, supporting various heterogeneous deployment scenarios, with multi-standard and multi-vendor support. As a result, it becomes a challenge for large-scale Telecom software companies to develop and test software for all deployment scenarios.


To address these challenges, we propose a framework for Automated Test Generation for large-scale Telecom Software systems. We begin by generating \textit{Test Case Input data} for test scenarios observed using a time-series Generative model trained on historical Telecom Network data during field trials. Additionally, the time-series Generative model helps in preserving the privacy of Telecom data. The generated time-series software performance data are then utilized with test descriptions written in natural language to generate \textit{Test Script} using the Generative Large Language Model. Our comprehensive experiments on public datasets and Telecom datasets obtained from operational Telecom Networks demonstrate that the framework can effectively generate comprehensive test case data input and useful test code.

\end{abstract}

\bstctlcite{IEEEexample:BSTcontrol}

\section{Introduction}

The advent of new 5G network technologies, the rising number of Radio Nodes in heterogeneous deployment scenarios, and the introduction of Open RAN (O-RAN) \cite{polese2023understanding} are challenges that Telecom vendors and mobile operators face to effectively test SW for myriad multi-vendor network deployment scenarios and various products. 

Software Testing (ST) is an important step in the SW development lifecycle to ensure service reliability of telecom systems. ST involves assessing the behavior of SW systems under test scenarios to detect failure cases.
Before rolling out any SW release in the network, each SW typically undergoes a rigorous testing process in a controlled environment. As mobile networks continuously evolve, Radio Nodes in operational networks can experience a problem due to new changes that can be hard to replicate in a controlled environment. The problem is exacerbated in 5G networks and beyond, where new SW features and configurations are being introduced at a higher pace.
The process involves creating many test cases to cover all potential test scenarios, and each one is crafted to validate a dedicated feature in the system. When developing a new SW version, the test cases are required to be executed and passed successfully. Once SW passes the review and test process, the new SW package can be released and potentially deployed in telecom networks. However, such test scenarios are crafted around the assumptions of network engineers, and new faults can emerge from uncovered scenarios.
Crafting and generating test cases is tedious, as it commonly requires manual work and deep technical expertise on potential deployment scenarios. Traditionally, testers define and execute a set of test case specifications. Each test case specification outlines test conditions, expected network behavior, and evaluation criteria and may involve devices or elements in the Telecom network.
To test SW behavior in the real world, mobile operators usually conduct field trials to determine the impacts of new SW capabilities. The new SW evaluation process includes executing a trial and collecting relevant data about resource usage and network performance. The process requires manual intervention from network engineers to assess the impact of the SW upgrade and analyze multiple inter-dependent interactions of base station SW and HW modules. Efficient ST can reduce Telecom Network Troubleshooting efforts \cite{FaultDetectionGNN}. Moreover, Telecom Networks consist of HW and SW components, so detecting faults early in the ST phase can potentially reduce site visit activities (e.g., climbing towers to check HW faults, etc). 

One of the challenges in the ST process arises in the gap between crafting test cases based on domain-expert assumptions and the performance of SW modules in the real world. As mobile networks continuously evolve, generating new test scenarios becomes harder due to the changes in the underlying network usage patterns or to unseen environmental conditions that can be hard to replicate in a controlled environment. 
Moreover, ST for Telecom SW Systems will be more challenging in O-RAN Networks \cite{tang2023ai}, where multi-vendor support is required for myriad RAN products and a wide telecom ecosystem. Ensuring robust and comprehensive testing and integration for multi\-vendor interoperability in O-RAN  \cite{tang2023ai} requires new ways to overcome the problem. Our method addresses some of these challenges and other challenges for ST.

There have been many efforts to automate ST Tasks. However, most ST methods are crafted for specific tasks and domains, trained on small-scale datasets, and typically require extensive domain expertise and task-specific model designs. This stands in stark contrast to \textit{Generative models} and \textit{Large Language Models} (LLMs) that can perform different sets of tasks.

In recent years, Generative models and LLMs demonstrated remarkable progress in different domains. While pre-trained LLMs have made impressive strides in different areas, including Telecom domain, their use in Telecom ST remains limited. In this work, we address this gap. We propose a framework to generate Test scripts using Generative Models. The core idea is to first generate the input time series into test case inputs and then combine generated data with natural language prompts to generate code using LLMs.

The main contributions of this work are summarized as follows:
    \begin{enumerate}
        \item We introduce a new framework to generate Test scripts using a hybrid generative model. Such a framework reduces lead time in Telecom SW development.
        \item We show the ability to leverage historical product performance data in an operational network to generate synthetic and expressive \textit{test case input data} in a privacy-preserving manner that covers testing scenarios. It is particularly important for O-RAN as multi-vendor support is required. 
        \item Our comprehensive evaluation using public and Telecom-specific datasets demonstrates the proposed method's robustness and generalization. 
    \end{enumerate}


\section{Related Work}

\subsection{AI for Test Generation}

In recent years, there has been a growing body of research \cite{anand2013orchestrated, harman2015achievements} that focuses on applying machine learning (ML) to automate the process of automated testing techniques. Several techniques have been used. For instance, Reinforcement Learning (RL) was employed in several studies \cite{abo2023role} to optimize Test Case Generation. Recently, LLMs started to be adopted for Software Engineering Tasks. The survey \cite{fan2023large} highlights recent techniques and discusses the limitations of using LLMs in this space. Alshahwan et al. \cite{alshahwan2024automated}  introduced TestGen-LLM that uses LLMs
to automatically improve existing human-written tests. TestGen-LLM was used to extend Kotlin
test classes at Meta for the Instagram and Facebook platforms.


 


In the context of the Telecom domain, Alzahraa \cite{salman2020test} conducted a study for Test Case Generation using Test Case Specifications. 
The method uses a multi-label text classification model trained on a corpus of test scripts and then classifies which test script should be used. The major drawback of the study revolves around the fact that test scripts should be provided, and unseen test scripts could not be suggested.

\subsection{Synthetic Data Generation}
Research in synthetic data generation has increased steadily. Notably, TimeGAN by Yoon et al. \cite{timegan} is one of the first models to maintain the temporal dynamics of the given data for synthetic TS data generation. The TimeGAN consists of four network components: an embedding function, recovery function, sequence generator, and sequence discriminator, which are trained by supervised and unsupervised losses \cite{timegan}. 

With the introduction of denoising diffusion probabilistic models (DDPM) by Ho et al. \cite{denoisingdiffusion}, plenty of generative models that utilize DDPM have been proposed recently. One of such is the Structured State Space Sequence Diffusion Model (SSSDS4) proposed by Miguel et al. \cite{SSSD}, which was used for TS imputation and forecasting. The network architecture uses a Structured State Space Sequence (S4) to sequence and model temporal dynamics. 





\subsection{Code Generation}

There have been several studies employing LLMs for SW Engineering tasks \cite{hou2023large} such as Code Generation \cite{zhang2023unifying}.

Ni et al. \cite{L2CEval} presented L2CEval, a systematic evaluation of the language-to-code generation capabilities of LLMs. Seven tasks were tested, including semantic parsing, math reasoning, Python programming, detection of factors that can potentially affect their performance based on model size, pretraining data, instruction tuning, and different prompting methods \cite{L2CEval}. Authors also measured confidence calibration for the models and conducted human evaluations of the output programs \cite{L2CEval}. 

Xion et al. \cite{programtest} explored the ability of LLMs to test programs/code by performing thorough analyses of recent LLMs for code
in program testing, they show a series of intriguing properties of these models and
demonstrate how the program testing ability of LLMs can be improved \cite{programtest}. Xio et al. \cite{programtest} further utilize their findings in improving the quality of the synthesized programs and show +11.77\% and +4.22\% higher code pass rates on the HumanEval+ dataset when comparing with the GPT-3.5-turbo baseline and the recent state-of-the-art.

In the context of Tests, Luis Perez et al. \cite{perez2021automatic} have proposed a model using a pre-trained language model (GPT2) to generate Python code. To train the model, they used a CodeSearchNet dataset consisting of +2M (comment, code) pairs with a mixture of different programming languages. Zheng et al. in \cite{zheng2023towards} discussed the challenges of using LLMs in SW engineering tasks.

While machine learning techniques have been used to automate ST Tasks that range from data input generation, test scenario generation, test scenario optimization, and test code generation, there is a lack of research papers that study the end-to-end flow. Our paper attempts to address this gap using hybrid generative models.
\section{Data}
\label{sec:dataCollection}


\subsection{\textbf{Telecom SW Time-Series Dataset}}
Telecom SW performance data was collected from a data lake accessible to the SW/HW diagnostics R\&D team. Multiple representative sites are selected from an operational network. SW Performance data may comprise one or more internal counters, and time series data points used to monitor SW Network performance, which is continuously from representative RAN nodes. The configuration data are relatively static and may comprise one or more configuration parameters and SW configuration parameters (e.g., frequency, HW technology, HW products, etc.).

\subsection{\textbf{Test Case Datasets}}
Due to the sensitive nature of company-specific code datasets, we use \textit{Publicly available datasets} to enable the reproducibility and reliability of our results. We use \textit{Company-specific test code dataset} to qualitatively assess the results.
\subsubsection{\textbf{Public Code Generation Datasets}}
\label{sec: Public Code Generation Datasets} 
We use the following datasets tailored for code generation tasks. 
\begin{itemize}
    \item \textbf{MBPP}  Mostly Basic Programming Problems (MBPP) dataset \cite{programsynthesisLLM} consists of 974 short Python functions designed to be solved by entry-level programmers, text descriptions of those programs, and test cases to check for functional correctness. This paper uses the sanitized version of the dataset, which consists of 257 samples.
    \item \textbf{Humaneval-X} \cite{humanevalx} consists of 820 hand-written problem-solution pairs, 164 problems for each of five languages. In this paper, we limit the evaluation to the Python language only.
\label{ref:publicdatasets}
\end{itemize}  

\subsubsection{\textbf{Telecom SW Vendor Test Case Specifications}}
Telecom SW Systems uses priority management tools to manage the SW Testing process. The system tracks a set of SW features, its functionality description, and corresponding Test specifications. Test specifications are documents that describe the test requirements and instructions for a SW system under test. Test specifications can be written in various formats, such as natural language or graphical representations. They may include testbed setup details, pre/post conditions, test procedure, and the criteria for passing or failing the test. For example, O-RAN outlines Test specifications \cite{tang2023ai} in the Test Specification portal.
In our work, we aim to use descriptions written in natural language in each test as an input for our method.

\section{Method}

\subsection{Problem Definition}

The main goal of this paper is to provide an end-to-end automated testing pipeline for large-scale Telecom SW products to (1) generate \textit{Test Case Input data} for test scenarios observed during field trials and (2) generate \textit{Test Script} for a given test description written in natural language.

Given (1) Telecom SW products performance data and (2) test description written in natural language, the pipeline should generate a test script with test cases.

\subsection{Method Overview}
\label{sec:MethodOverview}
In this section, we describe a Test Script Generation system for Telecom SW Systems. Algorithm \ref{alg:main_method_steps} summarizes the detailed steps.


Our method comprises three stages: 

\textbf{\textit{Stage I: Synthetic Test Input Data Generation}} 
When evaluating a system under test, test cases specify the conditions that must be executed to detect a failure in the system. A test case embodies the input values, which typically vary in nature, depending on the testing scenarios. To generate data input for test cases, we first train an end-to-end generative model that is able to capture the underlying distribution of each data cluster, as well as generate meaningful samples that belong anywhere in between the observed clusters. Such a model can extrapolate meaningful test cases in scenarios that are not present in the acquired data, yet still are plausible and reasonable given the configurations of network sites. As the data modality of interest is TS in nature, we propose the utilization of time-series generative models \cite{lin2023diffusion} such as Diffusion Models with Structured State Space Sequence (S4) and timeGAN. Such models not only are very expressive in regard to the distributions they can describe but also allow a high degree of control over what type of data we wish them to generate. For example, we can specify if we wish to generate around a cluster found from our real data or something that lies in between, all while respecting the underlying data distribution.

\begin{algorithm}
\begin{algorithmic}[1]
\STATE Conduct field trial and obtain Telecom SW performance data from a representative set of RAN nodes in Telecom network $NW_{operator}$. \\ 
\STATE For each RAN node $R_{i}$ in Telecom network $NW_{operator}$, we consider a multivariate time series of $K$ features with the TS-length of $L$:
\[(\mathbf{x}^l)_{l=1}^{L} = (x_1^l, \ldots, x_K^l)_{l=1}^{L}\]
where each data point $x_k^l\in \mathbb{R}$ represent a Telecom SW performance value that is collected at timestamp $l$ for the SW feature $k$.
\STATE Produce synthetic data $\hat{\mathbf{x}}_{0}^{l}$ using a generative model. \\ 
\STATE Given a set of \textit{Test Problems} $TP \in \mathbb{N}^+$ Provide Test Instruction $D$ for each \textit{Test Problem} ${i}$.
\STATE For each {Test Instruction} $D_i$, generate Test cases $M_i$ from Test Inputs $\hat{\mathbf{x}}^{l}$, where $M_i \in \mathbb{N}^+$, and return a string output $O_i$.
\STATE Generate a prompt $P_i$ = concat($D_i$,$O_i$).
\STATE Feed the prompt $P_i$ to LLM trained on Code corpus. \\ 
\end{algorithmic}
\caption{Main steps of the method}
\label{alg:main_method_steps}
\end{algorithm}

\textbf{\textit{Stage II: Test Case Script Generation}} Synthetic data generated from the previous end-to-end model is then extracted and fed as examples within a prompt for an LLM trained on Code. Such models are known to be able to generate meaningful code, based on a brief description of the task, accompanied by a few example cases. The utilization of such a complex model allows us to alleviate the traditional limitations tied to fitting the synthetic data onto predefined, fixed templates.

Overall, our solution allows us to generate high-quality, controllable synthetic data which we can then leverage to generate completely novel test cases.

\begin{figure}[ht!]
     \centering
     \includegraphics[scale=0.2]{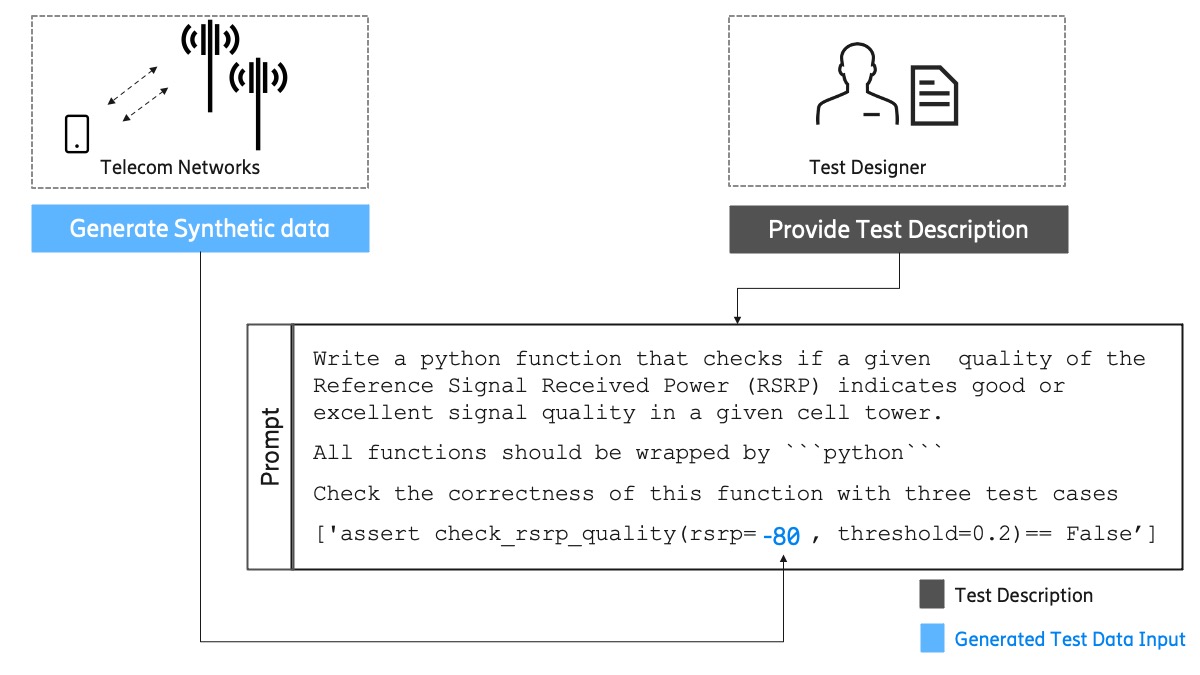}
     \caption{Prompt Encoding.}
     \label{fig:prompt_encoding}
     {\addvspace{-1\baselineskip}}
 \end{figure}

\textbf{\textit{Stage III: Code Generation for Telecom SW Testing}} 
Generate a Test script using a Prompt (human language that describes the intention behind the test) and examples of synthetic data. Then, The LLM is then used to generate a test script and test case using the before-mentioned prompt.
Figure \ref{fig:prompt_encoding} depicts an example of a prompt with synthetic data.

\subsection{Evaluation Metrics}
 In the synthetic data generation task, we applied t-SNE. For all measurements of the t-SNE, the perplexity value is assigned to $100$. The t-SNE is intended to be used as an evaluation metric by applying the original and generated dataset and visualizing the results by a 2-dimensional plot.
 In the code generation task,
 the assessment is used to evaluate the test cases generated by our method.
The process builds upon insights described in \cite{hu2022correlating} to show the content and the effectiveness of the output of the LLM models. There have been proposals for other evaluation metrics that compare the text output such as codeBLEU, METEOR, and ChrF \cite{LLMeval}. In our work, we want to evaluate the usability of the generated code based on industrial needs. We favored the definition of the following three metrics:

\textbf{Meteor}: \cite{LLMeval} compares machine-produced content with human-produced content.

\textbf{Content Score (CS)}:
 \begin{equation}
     CS = \frac{1}{N}{\sum_{i=1}^{N}o_{i}}
 \label{eq: CS}    
 \end{equation}
 where $o_{i} \in \{0, 1\}$ where $o_i = 1$ means that the model yielded an output, and Python successfully executed the output function and $o_i = 0$ means that the executed failed, and $N$ is the number of total test samples (defined in \ref{sec: Public Code Generation Datasets}).

\textbf{Test Case Effectiveness Rate (TCER)}:
 \begin{equation}
    TCER = \frac{1}{N}\sum_{i=1}^{N}\left[\frac{\sum_{j=1}^{M_i}{c_j}}{M_i}\right]
 \label{eq: TCER}
 \end{equation}
 where $M_i$ is the number of test cases for test sample $i$, and $c_j \in \{0, 1\}$ where $c_j = 1$ means that the output function passed for test case $j$, and $c_j = 0$ means that the output function failed for test case $j$.

 \begin{figure}[ht!]
     \centering
     \includegraphics[scale=0.2]{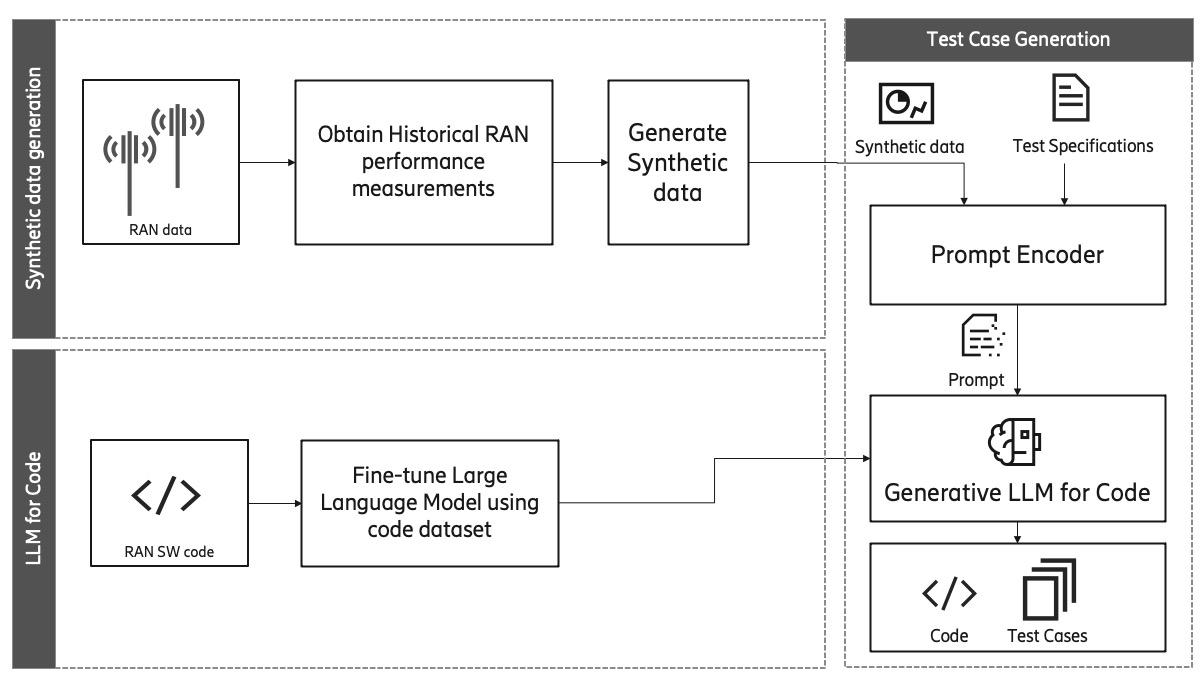}
     \caption{Main Components of the Framework.}
     \label{fig:components}
 \end{figure}

\section{Experiments}
In this section, we present and discuss our findings. We begin by examining the outcomes of Synthetic Test Data Input generation on telecom dataset. Next, we evaluate and compare the performance of LLM models on Test Code Generation.

\subsection{Synthetic Data Generation}

For Synthetic Test Data Input generation Test, we evaluate the following state-of-the-art models for multivariate TS data generation, SSSDS4 \cite{SSSD} and TimeGAN \cite{timegan}.

 \begin{figure}[!h]
     \centering
     \includegraphics[scale=0.5]{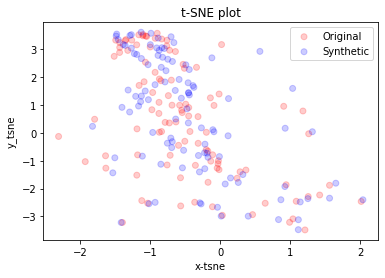}
     \caption{SSSDS4 t-SNE plot on telecom data.}
     \label{fig:SSSDS4 tsne newsite}
     {\addvspace{-1\baselineskip}}
 \end{figure}
 
 \begin{figure}[!h]
     \centering
     \includegraphics[scale=0.5]{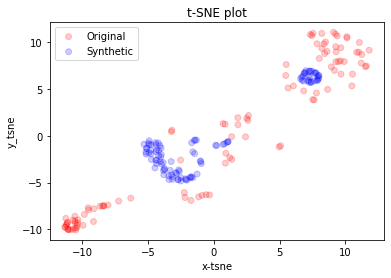}
     \caption{timeGAN t-SNE plot on telecom data.}
     \label{fig:tsne_newsite_timeGAN}
 \end{figure}
The results of the t-SNE plots can be seen in Figure \ref{fig:SSSDS4 tsne newsite} and Figure \ref{fig:tsne_newsite_timeGAN}. Both generative models follow the distribution of the original data in the 2-dimensional t-SNE plot well. The timeGAN follows and is clustered based on the pattern of the original data points whereas the SSSDS4 is scattered wherever the original data points may exist.
 
\subsection{Code Generation}
For Test Code generation, we evaluate the following state-of-the-art LLMs for code generation: \\
1) Mistral-7B \cite{jiang2023mistral} (General LLM) trained on the general corpus. The model was trained with a focus on efficiency to enable applications on real world. \\
2) CodeLLama \cite{roziere2023code} (Specialized LLM) trained on code corpus, with a focus on code generation tasks.

\begin{figure}[!h]
     \centering
     \includegraphics[scale=0.78]{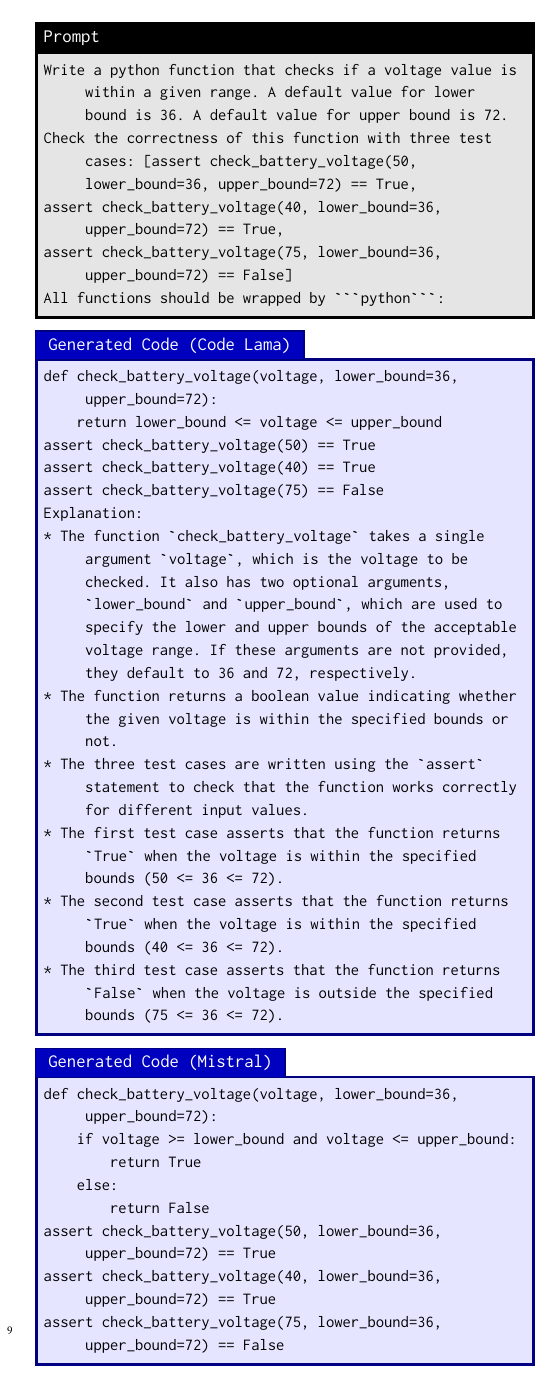}
     \caption{Example of generated code using CodeLLama and Mistral on Telecom samples.}
     \label{fig:code_example}
 \end{figure}

The code generation method was assessed and quantified using two public datasets MBPP \ref{ref:publicdatasets} and Humaneval-x datasets \ref{ref:publicdatasets} as described in \ref{sec:dataCollection}. Our experiments demonstrated the following:
\begin{itemize}

\item \textbf{Improved Content Generation using Specialized LLMs:} The use of CodeLLama, which is trained on code corpus, led to significant improvement in code content generation (see Table \ref{tab:quantitative results}). The model can also generate an explanation along the code, The middle code box of Figure \ref{fig:code_example} depicts an example for a given test problem from CodeLLama. CodeLLama was able to outperform Mistral in generating better Code format (output content), the same insights are shown in recent works \cite{wei2023magicoder} \cite{yang2024llm}  where authors show that LLMs that are trained on better instruction data lead to better code models.
\item \textbf{Similar Effective Performance:} Even though Mistral was not trained solely on code corpus, it showed similar effective performance and demonstrated good reasoning capabilities in terms of TCER \ref{eq: TCER}. Both of the models were able to reason and generate test code. 
\item \textbf{The choice of metrics matters:} Code generation using LLM remains a complex task. The model output is not always formatted to the dataset's desired format, which does not necessarily mean that the output of the model is inadequate. Potentially it means that the CS metric may not always effectively display the model's performance output when the CS score is low as it could have produced the right function but unnecessary details came along with the output, this was not an issue for CodeLLama but can be a reason behind the low CS \ref{eq: CS} results of the Mistral. Moreover, recent benchmarks \cite{honarvar2024turbulence} highlighted code correctness and robustness issues.
     
\end{itemize}
Furthermore, we conducted a qualitative analysis of the internal Telecom dataset. Due to the sensitivity of the propriety code content, we used public dataset to show the effectiveness of our approach. Figure \ref{fig:code_example} shows an example in a Telecom context using both LLMs for code models. In addition to previous insight, we observed that different flavors of \textit{LLMs for Code} are sensitive to the prompt format. The observation has been highlighted in an extensive evaluation of LLMs for Code \cite{L2CEval}. Additionally, recent works \cite{ridnik2024code} suggest exploring ways beyond traditional prompting. Moreover, while there are other efficient open-access proprietary LLMs, privacy and security guidelines in the Telecom domain limit real-world applications.

\begin{table}[!ht]
    \centering
    \caption{The results of the LLMs on the code datasets.}
    \label{tab:quantitative results}
    \begin{tabular}{|c|c|c|c|c|}
         \hline
         \textbf{Model} & \textbf{Dataset} & \textbf{CS} & \textbf{TCER} & \textbf{Meteor}  \\
         \hline
         Mistral-7B & MBPP & 63.8\% & 33.6\% & 26.4\% \\
         \hline
         Mistral-7B & Humaneval-x & 42.7\% & 37.2\% 
         & 64.0\%\\
         \hline
         CodeLLama-7B & MBPP & 95.1\% & 38.1\% &
         27.2\%\\
         \hline
         CodeLLama-7B & Humaneval-x & 84.0\% & 1.9\% & 22.7\% \\
         \hline
    \end{tabular}
    
    {\addvspace{-1\baselineskip}}
\end{table}

\section{Conclusion}

The paper presented a framework to generate a test script for Telecom SW applications using a two-stage model. In the first stage, we used field trials SW performance data to generate test input. In the second stage, we use the test data input with the test description expressed in natural language to generate the script. The paper shows promise in adapting LLMs for automated test generation in Telecom. 

In the first stage, we conducted experiments to validate the performance of the generative models in producing synthetic data that mirrors the characteristics of the original telecom dataset. 
Results suggest that the applied TS generative models effectively generate synthetic data. The synthetic data aligns with the distribution of the original dataset when visualized in a 2-dimensional t-SNE space. 
In the second stage, we conducted experiments to validate the ability of LLMs to generate code and give test descriptions. Our experiments on public \textit{Code Generation datasets} provided good results. Furthermore, we conducted a qualitative analysis using test descriptions from the Telecom domain. 
Further research should explore possibilities to explore multi-modal models to feed the joint data modalities (i.e., Telecom Network TS Performance data and natural description) to LLM. 
This work is considered one of the early efforts that used LLMs and GenAI for Telecom SW Testing. We hope our paper can provide useful insights for the \textit{TelcoAI} community into the adoption of LLMs in Telecom.




\bibliographystyle{ieeetr}
\bibliography{bibliography}
\end{document}